# Open Information: A Defining Perspective on Web Datasets for Carbon Pricing

Sidharth Mallik
*Centre of Excellence for Data Science, Artificial Intelligence and Modelling*
*University of Hull*
Hull, UK
orcid.org/0000-0003-2562-1738

Anastasios Megaritis
*Hull University Business School*
*University of Hull*
Hull, UK
orcid.org/ 0000-0002-6797-6636

Waymond Rodgers
*Woody L. Hunt College of Business*
*University of Texas at El Paso*
El Paso, USA
orcid.org/0000-0003-4349-5667

***Abstract*—The impact of web datasets on market prices has suggested the development of new sources of information, such as social media and web portals, indicating the possibility of an emergent phenomenon. We propose a defining perspective, termed *open information*, that adds to the existing types of *public* and *private* information. We demonstrate their existence and justify material significance for pricing. In this respect, we present statistical hypotheses to test for a web dataset, GDELT, integrated for carbon pricing, that is represented by EU Allowance spot prices. Tests are designed with VAR and GARCH-X formulations, and return forecasting. The outcomes cannot rule out the material existence of open information. The result is significant for providing a conceptual basis to integrate a vast number of web datasets as alternative data in investment decisions[1].**

## I. INTRODUCTION

Identifying information sources as *public* or *private* is a cornerstone of *neoclassical finance* [1, 2]. Major events in financial markets, namely, the Twitter-driven[2] *flash crash* on April 23, 2013, the speed of *bank run* of the Silicon Valley Bank (SVB)[3], and price moves in stocks of GameStop[4], have indicated the possibility of an emergent phenomenon in modern markets. We question if the introduced categorisation of information sources is exhaustive in light of such modern developments, prominently from two major innovations, the birth of the *internet* and the proliferation of datasets in society referred to as the phenomenon of *datification* [3]. A vast number of web datasets is available from third-party web platforms [4]. A significant body of literature exists to demonstrate their impact on market pricing, in the capacity of new datasets arising out of sources like social media, viral headlines, or influencer marketing. Consequently, we perceive web datasets as a new source of information. Termed as Open Information (OI), we therefore propose a defining perspective in DEFINITION 1.

> Open Information is the information content of a web dataset that is widely available and is not a designated form of public information.

*DEFINITION 1: DEFINING PERSPECTIVE ON WEB DATASETS*

From the definition, any form of public information (PI) is excluded, and this also differentiates from private information citing wide availability, a characteristic of the web's open architecture. The differentiation goes beyond the perspective and TABLE 1 identifies characteristics of the three types of information.

TABLE 1: CHARACTERISTICS OF PUBLIC, PRIVATE AND OPEN INFORMATION

| Characteristics | Public Information | Private Information | Open Information |
|---|---|---|---|
| **Availability** | Online or offline | Private networks | Web datasets |
| **Regulated** | Yes | Yes | Monitored for impact |
| **Description** | Firm sensitive | Firm sensitive | Pricing sensitive |
| **Examples** | Annual reports, Mergers & acquisitions | Strategic information on Board decisions, Career moves of executives | Viral headlines on the sector, posts on foreign competitors' breakthroughs, user responses to |

[1] Material presented here must not be construed as investment advice and the reader is expected to perform own due diligence before applying the outcomes.
[2] Twitter is a former name of a social media platform that has undergone restructuring since the event occurred.
[3] SVB was a prominent bank in the USA, and their assets were restructured after the bank run with several divisions now existing as part of the inorganic restructuring.
[4] GameStop is a retailer.

| | | | |
|---|---|---|---|
| | | | posts from executives, and influencer activities |
| **Consumers** | Stakeholders of the firm | Strategic stakeholders involved in a sensitive decision | Investors, investment algorithms, firm clientele |
| **Dissemination** | Periodic and controlled | On sensitive occasions and privately available | Widespread and open, possibly sudden and fast |
| **Jurisdiction** | Designated regulator inside the country | Designated regulator inside the country | Impact is over the web, which has worldwide connectivity |
| **Impact** | Fundamentals of the company | Impact related to an upcoming event or an immediate investment decision | Impact could be a price shock in an extreme situation or as an input to investment decisions, investor sentiment |
| **Origin** | Stakeholder related to the business | Stakeholder related to the business | Web platforms, specialised data sources |

We study the impact of OI on carbon pricing through tests for integrating a web dataset, GDELT[5], to the pricing of the EU Allowance (EUA) spot. After a brief literature review, the methodology highlights details on modelling OI, followed by tests from well-established time series approaches such as GARCH, VAR and return forecasting. The significance for investment decision-makers is highlighted through a decision model, the Throughput Model (TM) [5, 6]. We then discuss the implications of our findings and conclude with notes on existence and materiality. Our perspective and the outcomes are a discovery within an emergent phenomenon that contributes to explaining the functioning of modern markets, therefore holding significance for an array of stakeholders exploring alternative data in financial markets.

## II. LITERATURE REVIEW

There exist instances from literature demonstrating the impact of web datasets on pricing, especially arising from platforms like social media [7-11]. In addition to social media, there exist online portals like Yahoo Finance[6] with a significant following and thus having the potential to impact prices. Some researchers have suggested characterising such datasets as an information source [12, 13], therefore motivating new development. Their impact is prominent with certain market events introduced earlier. The Twitter-driven flash crash on April 23, 2013, was understood to have been caused by a hack on the account of Associated Press when a false tweet about explosions at the White House was reported, only to have been removed moments afterwards.[7]. The markets reacted briefly and corrected soon after. The second instance with the price movement of GameStop[8] has been studied before[9] [14], where a group of retail traders communicating over Reddit were able to raise the stock prices in spite of institutional traders holding an opposing position. The final example is that of the bank run of SVB, an event that occurred in around 48 hours, at a shocking pace considering the size of the bank. The Senate Committee that examined the bank failure discussed the possibility of the first-ever social media-led bank run.[10].

Having motivated a background, we introduce the literature reviewed for the specific aspects of our approach, which we have taken to demonstrate the existence of OI and justify their material significance for investment decisions. In this respect, we consider one of the modern markets, that of carbon pricing with EU Allowances as part of the EU ETS. Developing over the last two decades, carbon pricing has significance for estimating the social cost of carbon [15] and has implications for the energy transition [16, 17]. However, questions have been raised towards their form, prominently on the possibility of a potentially new asset class [18, 19] and their pricing characteristics such as in relation to market efficiency [20, 21]. Therefore, finding answers to such questions is a relevant area for new research.

[5] Available at https://www.gdeltproject.org/.

[6] According to SEMRUSH, which provides traffic data on websites, Yahoo Finance received over 140 million visits in the last month alone (Source: https://www.semrush.com/website/finance.yahoo.com/overview/) (Accessed 29-04-2026).

[7] A CNBC account of the event is available at https://www.cnbc.com/2013/04/23/false-rumor-of-explosion-at-white-house-causes-stocks-to-briefly-plunge-ap-confirms-its-twitter-feed-was-hacked.html (Accessed 20-01-2026).

[8] Stock prices on NASDAQ are available at GameStop Corporation Common Stock (GME) Advanced Charting | Nasdaq (Accessed 20-01-2026). The period towards the end of 2020 and the beginning of 2021 is being referred to.

[9] An interesting documentary on the topic is called *GameStop: Rise of the Players*.

[10] The hearing video is available at https://www.banking.senate.gov/hearings/examining-the-failures-of-silicon-valley-bank-and-signature-bank.

To include the component of OI, we consider GDELT, a web dataset that integrates news headlines worldwide. While our study with the dataset is one of the earliest, there have been instances where GDELT has been researched for pricing [22]. To demonstrate the existence of OI through statistical hypothesis testing, we consider two prominent approaches, GARCH [23] models and their extension to include an external regressor as the GARCH-X formulation [24], and with Vector AutoRegressive (VAR) models. In addition, results on tests for returns forecasting are provided. The corresponding investment decision is modelled with the TM [25] that provides a 2-stage decision framework for integrating information and has been previously applied to model management decisions. Furthermore, we have taken a Data Science [26] approach to modelling where tests are performed *in-sample* and *out-of-sample*, with a *window* for lookback data and *leave-one-out* validation [27]. Such a procedure makes the approach statistically robust to unseen data and is contemporary in modelling.

## III. METHODOLOGY

Our approach hinges upon the realisation that a new information source, in form of a web dataset, can explain the structure of a price time series and can be validated statistically. The choice of carbon pricing for tests is justified on the grounds that such markets are modern, there is research to suggest informational inefficiency, and there exist datasets relevant to test our hypotheses. Carbon prices are represented by EUA spot prices. We consider the logarithmic returns series defined in (1).

$$r_t=100*\left(\log(p_t)-\log\left(p_{(t-1)}\right)\right) \qquad (1)$$

Headlines provided by GDELT are taken as a proxy for information. GDELT categorises news through provided codes, where environmentally relevant headlines are identified by “ENV” in the Secondary Role Code. Moreover, the headlines are categorised based on their origin, where some are identified as sources for PI, while the remaining are categorised as those for OI, summarised in TABLE 2.

TABLE 2: GDELT CAMEO CODES TO CATEGORISE ENVIRONMENTAL NEWS HEADLINES AS PUBLIC OR OPEN INFORMATION

| CODE | Description | Type of Information |
|---|---|---|
| BUS | Business: businesspeople, companies, and enterprises, not including MNCs | Public |
| GOV | Government: the executive, governing parties, coalition partners, and executive divisions | Public or Open |
| MNC | Multi-national corporations | Public |
| Others | Various descriptions | Open |

The categorisation proves to be an advantage with GDELT over alternative sources, such as social media datasets or those from online portals, in the sense that this avoids the ambiguity arising from keyword selection to categorise news headlines. Furthermore, the differentiation between PI and OI is identified immediately from their definitions.

The next step is transforming the data for modelling. Textual data as pricing input is well established in the literature [28]. A way to transform such data for modelling is to take the count of news headlines as a source of information [29]. To differentiate the count series from the two types of sources, we introduce a notation, $P$ for PI and $O$ for OI. Furthermore, we are alert to the possibility of them being correlated, highlighted as PROPOSITION 1.

$$\text{Correlation}(P, O)\neq 0$$

*PROPOSITION 1: OI AND PI ARE CORRELATED*

To capture this correlation through a variable transformation, we introduce the formulation in (2).

$$\log(O_c)=\phi*\log(P)+\log(O_u) \qquad (2)$$

The equation first transforms the time series from whole numbers to real numbers through a log transformation, and then separates two components of OI based on their correlation with $P$, the correlated $O_c$ and the uncorrelated $O_u$. Furthermore, the constant term is not included on the grounds that both forms of information arise at the occurrence of the same set of events. Therefore, each term increases or decreases simultaneously and their scales are related by the respective coefficients. The formulation also enables defining key qualities of *existence* and *materiality* through PROPOSITION 2.

OI exists if $O_c$ is shown to exist, while OI is materially significant if $O_u$ presents information over and above $P$.

*PROPOSITION 2: EXISTENCE AND MATERIALITY OF OI*

The importance of the proposition is that even in the presence of a correlation, when the uncorrelated component has an impact on pricing, OI is materially significant. This rules out the possibility that the information content in OI is merely a spillover from PI. Furthermore, the formulation enables numerically estimating the uncorrelated component by *regressing out* PI. The residuals from the regression constitute the desired time series.

## IV. TESTS

TABLE 3: CONFIGURATION OF DATASETS

| Configuration Header | Value |
|---|---|
| Start date | 01-04-2013 |
| End date | 31-12-2023 |
| Source of GDELT | GDELT Events Database |
| Size of total data | Order of 100 GB |
| Source of EUA spot prices | The International Carbon Action Partnership[11] |
| Price series | End-of-day Close |

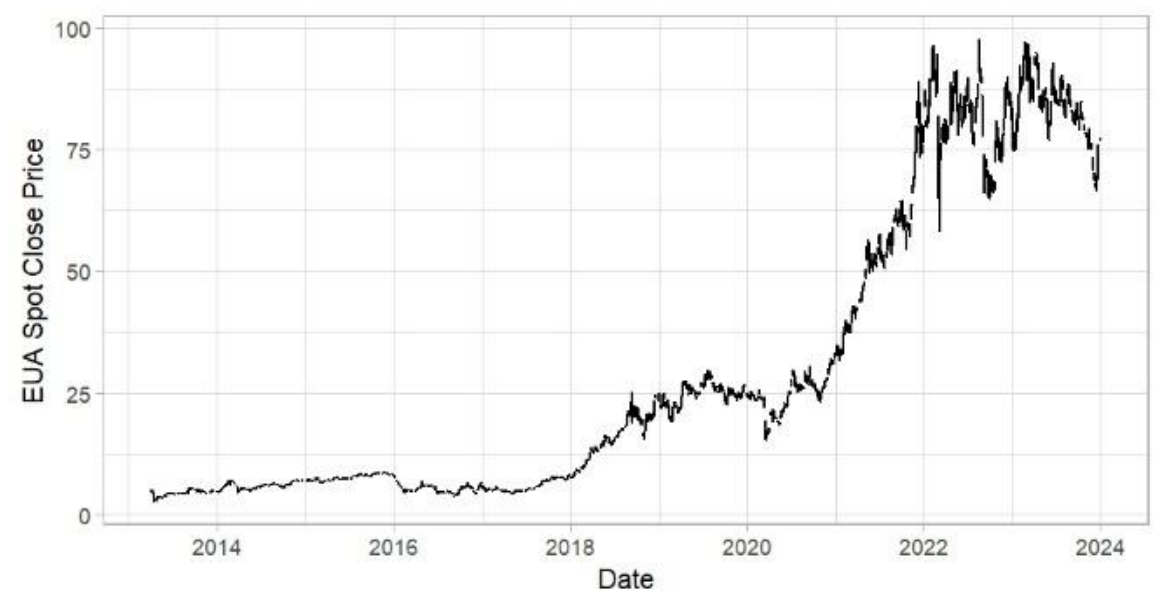

*Fig 1: Carbon Prices Represented by EUA Spot Close Prices*

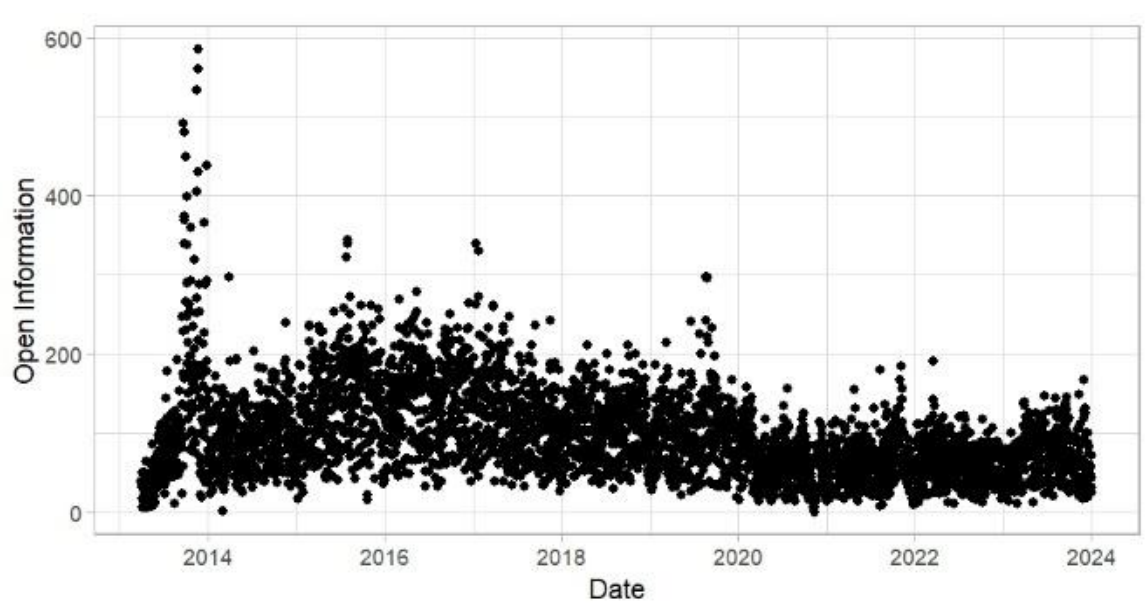

*Fig 2: Open Information Represented by Count of Selected Headlines*

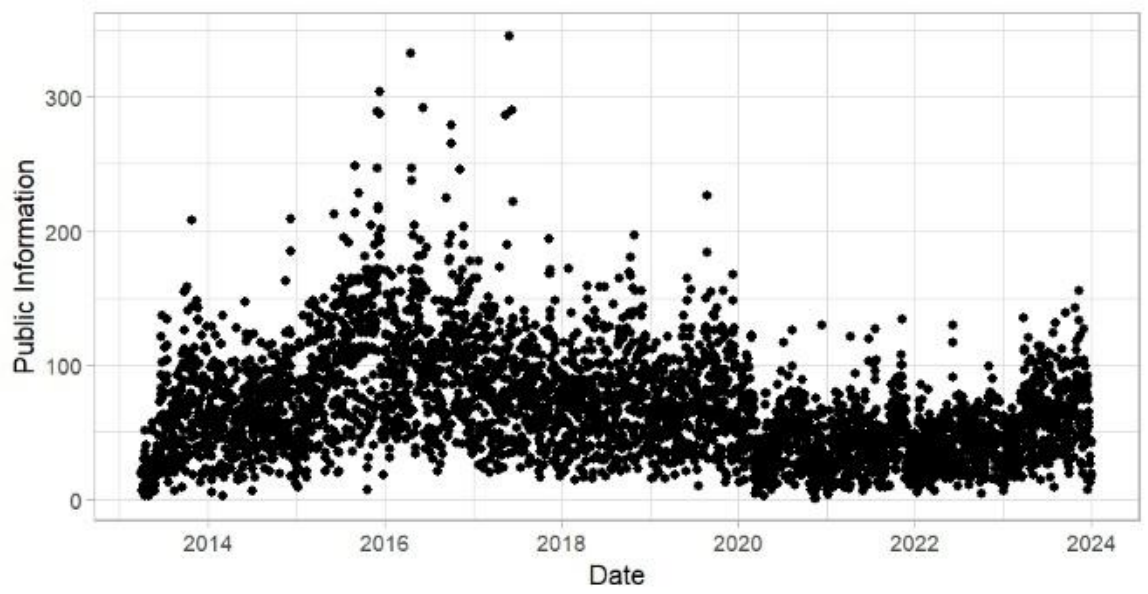

*Fig 3: Public Information Represented by Count of Selected Headlines*

[11] The website for ICAP is https://icapcarbonaction.com/en (Accessed 23-01-2026).

329315961 20131224 201312 2013 2013.9699
BUS
BUSINESS
BUS 0
042 042 04 1 1.9
10 1 10 5.21653543307087
0
2 Minnesota, United States
US USMN 45.7326 -93.9196 MN
2 Minnesota, United States US
USMN 45.7326 -93.9196 MN 20141224
http://www.prweb.com/releases/2014/12/prweb1241218
5.htm

*DEFINITION 2: SAMPLE ROW FROM GDELT*

The configuration of datasets for testing is summarised in TABLE 3, and a sample row from GDELT is exhibited in DEFINITION 2. The relevant time series obtained from GDELT corresponding to OI and PI are summarised in TABLE 4. The counts obtained as a result of the selection are exhibited in Fig 2 and Fig 3. Even though headlines categorised as GOV could produce both OI and PI, we have categorised them as PI citing difficulty in differentiating the sub-categories. Such a misclassification weakens OI. The weakening is acceptable on the grounds that a more accurate classification would improve the strength of OI. Therefore, if our tests are significant in the weaker form, the expected output is only stronger without the misclassification.

TABLE 4: TIME SERIES FOR OI AND PI

| Time Series | Categories | Notation |
|---|---|---|
| Public information | ENV, BUS or MNC or GOV | $P$ |
| Open information | ENV, (not BUS) and (not MNC) and (not GOV) | $O$ |

### A. *Test for PROPOSITION 1*

We report two sets of correlation, one between $O$ and $P$, and the other between their logarithmic transformations. The outcome is summarised in TABLE 5 and confirms the PROPOSITION 1.

TABLE 5: CORRELATION BETWEEN OI AND PI

| Time Series | Pearson's Correlation | Kendall's τ | Spearman's ρ |
|---|---|---|---|
| Correlation between $O$ and $P$ | 0.68 | 0.57 | 0.76 |
| Correlation between $log(O)$ and $log(P)$ | 0.76 | 0.57 | 0.76 |

## B. Extracting the Uncorrelated $O_u$

We extract $log(O_u)$ defined in (2). $O_c$ is the time series $O$, and therefore a linear regression of $log(O_c)$ versus $log(P)$ yields the desired time series as residuals. The regression outcome is summarised in TABLE 6.

TABLE 6: REGRESSION FROM (2) TO OBTAIN RESIDUALS AS LOG($O_C$), NOTING THE MISSING INTERCEPT TERM.

| | (1) |
|---|---|
| Dependent variable: | $log(O_u)$ |
| $log(P)$ | 1.08*** (0.002) |
| Observations | 2,803 |
| Adjusted R-squared | 0.992 |

Notes: Standard errors in parentheses. *** p<0.01, ** p<0.05, * p<0.10.

The combined time series has 106 missing values, resulting from missing dates and when the count is zero. A randomised bootstrap algorithm then imputes for the missing values. The time series thus obtained has a mild correlation to the magnitude of the returns from carbon prices, as reported in TABLE 7[12].

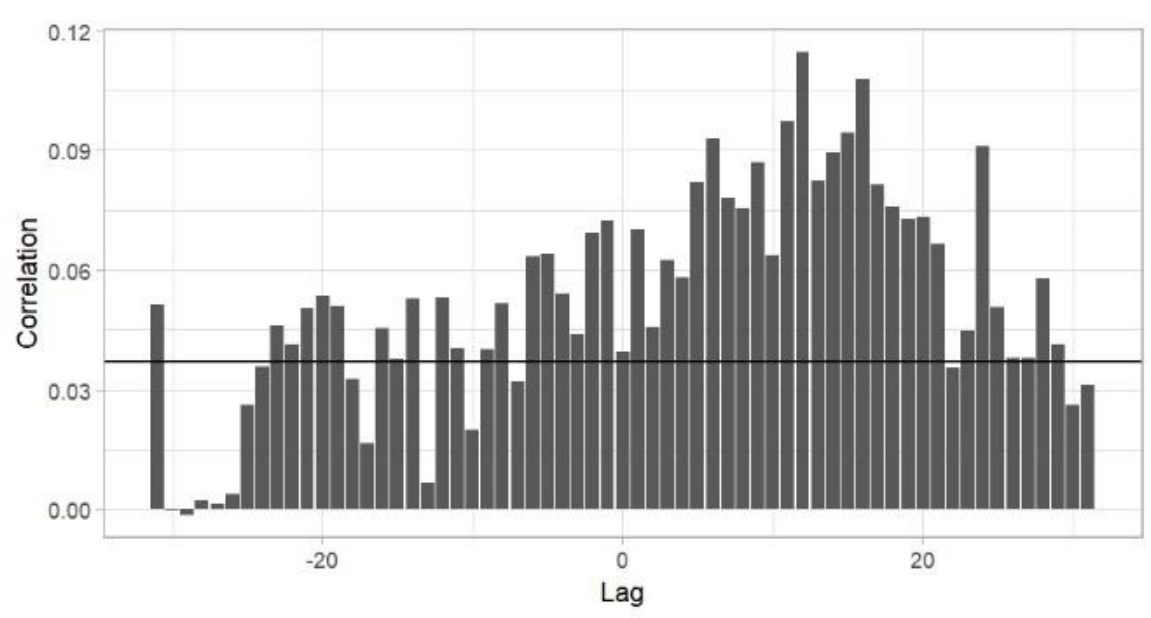

*Fig. 4: Cross-correlation Between $Log(O_{u,t})$ AND $|r_t|$ with the Lag k Indicating the Correlation Between $Log(O_{u,t+k})$ and $|r_t|$.*

In addition, we estimate the cross-correlation between $log(O_{u,t})$ and $|r_t|$ exhibited through Fig. 4. The correlations suggest that lagged values have an influence, thus motivating the question on the functional form for this influence, one we explore with tests in the subsections that follow.

TABLE 7: CORRELATION BETWEEN LOG($O_{u,t}$) AND $|r_t|$.

| **Correlations** | $log(O_{u,t})$ | $\|r_t\|$ |
|---|---|---|
| $log(O_{u,t})$ | 1 | 0.040 |
| $\|r_t\|$ | 0.040 | 1 |

## C. Test with a VAR Formulation

We consider a VAR model[13] in (3) and (4)[14].

$$\mathbf{y_t}=A_0+A_1\mathbf{y_{(t-1)}}+\ldots+A_p\mathbf{y_{(t-p)}}+\mathbf{u_t}, \quad (3)$$

$$\text{where } \mathbf{y_t}=\left(\log(O_{u,t}),|r_t|\right)^T \quad (4)$$

Results with the first lag only and a constant term[15] is presented in TABLE 8. More lags were tested and revealed similar results.

TABLE 8: RESULTS FROM THE VAR MODEL OF (3).

| | (1) | (2) |
|---|---|---|
| Dependent variable: | $log(O_{u,t})$ | $\|r_t\|$ |
| Lag 1: $log(O_{u,t-1})$ | 0.34*** (0.018) | 0.110** (0.052) |
| Lag 1: $\|r_{t-1}\|$ | 0.01** (0.004) | 0.18* (0.091) |
| Intercept | -0.01 (0.11) | 1.67*** (0.054) |
| Observations | 2,753 | 2,753 |
| Adjusted R-squared | 0.121 | 0.037 |
| Multiple R-squared | 0.121 | 0.038 |

Notes: Standard errors in parentheses. *** p<0.01, ** p<0.05, * p<0.10.

## D. Test with GARCH-X Formulations

We consider an AR(1)-X-GARCH formulation[16] in (5) and (6) with an extra term for $log(O_u)$[17].

$$r_t=\mu+\alpha_1 r_{(t-1)}+\beta_1 \log(O_u)_{(t-1)}+\eta_t \quad (5)$$

$$\text{and} \quad \sigma_t^2=\omega+\alpha_2\eta_{(t-1)}^2+\beta_2\sigma_{(t-1)}^2 \quad (6)$$

To perform model estimates, an approach with an *in-sample increasing window* of past returns was taken, starting from the first value. For the last value of the window, we started with the index 2501 and continued to the last index 2753, therefore creating 253 sampling windows. In each window, two formulations were estimated, the GARCH-X model defined earlier and the corresponding GARCH without the term for $log(O_u)$. To check for the improvement in modelling accuracy, two criteria, those of AIC and BIC were calculated for the formulations. Where both indicated an improvement, the impact and significance of the term was noted. In this subsection, we test for *in-sample* performance improvement only. The tests were extended to a set of GARCH variants, Exponentially Weighted Moving Average (EWMA), fractional-GARCH (fGARCH) and integrated-GARCH (iGARCH). The outcome is summarised in TABLE 9.

[12] Because of the randomised form of the imputation algorithm, exact values might vary from the displayed ones. However, the magnitude obtained with multiple estimates were of a similar order.

[13] In an alternative formulation with returns instead of absolute returns, the coefficient on $log(O_{u,t})$ was statistically insignificant.

[14] Bold notation represents vectors.

[15] A significant result was also obtained without an intercept.

[16] An alternative formulation with the added term in the equation for variance was also considered, where the coefficient was found to be statistically insignificant.

[17] Different statistical distributions for the $\eta$ term were considered, Normal and Student's t, with similar results.

TABLE 9: RESULTS ON IMPACT SIGNIFICANCE FROM AIC AND BIC IN GARCH MODEL VARIANTS

| Model Variant | Improvement in AIC (for a sample size of 253) with X term | Improvement in BIC (for a sample size of 253) with X term |
|---|---|---|
| **GARCH (1,1)** | 253 | 253 |
| **fGARCH** | 253 | 253 |
| **EWMA** | 248 | 0 |
| **iGARCH** | 0 | 0 |

### E. Tests on Returns Forecasting

In the same window configuration as the previous subsection, an *out-of-sample leave-one-out* approach was taken, where the prediction error for the immediately next value of the sample was considered. The window was then expanded to test for the next index as described earlier as an *increasing window*.[18]. Results are summarised in TABLE 10.

TABLE 10: PREDICTION ERROR IMPROVEMENT WITH EXTERNAL FACTOR[19]

| Metric | AR(1)-GARCH(1,1) | AR(1)-X-GARCH(1,1) | Result of Comparison (Improvement Order) |
|---|---|---|---|
| **Mean Error** | -0.05112 | -0.04284 | Improved by ~10% |
| **RMSE** | 1.94575 | 1.94543 | Improved by ~0.01% |
| **MAE** | 1.52465 | 1.52311 | Improved by ~0.1% |

## V. DISCUSSION

We have statistically tested for Propositions 1 and 2. While the outcome for PROPOSITION 1 from Tables 5 & 6 is clearly justifiable, those for PROPOSITION 2 in terms of material significance have been mixed. TABLE 8 demonstrates the significance of the lagged term $log(O_{u,t-1})$ in the equation for $|r_t|$ and therefore indicates success for our hypothesis test. However, the low explanatory power of the VAR model noted from $R^2$ values is a setback. To realise the importance, we compare the influence with the autocorrelation of $|r_t|$, which, from the studies on *volatility persistence*, is a well-documented significant characteristic of price returns. We note that the magnitude of $log(O_{u,t})$ is lower than $|r_t|$ to the order of 1 decimal point. Given that the coefficients of the two lagged terms are similar, this implies that the influence from $log(O_{u,t-1})$ is of the order of a tenth of that from volatility persistence. This indicates a significant yet feeble effect. Moreover, the results from TABLE 9 indicate that the impact varies depending on model selection, and we cannot rule out the possibility of the influence being overridden by another variable or a transformation as noted in the case of the integrated-GARCH model. In this sense, the result is sensitive to model selection. TABLE 10 corroborates this characteristic, where certain measures of error like *Mean Error* display greater improvement (~10%), while the remaining improve to a lesser extent. While interpreting the outcomes, two key points are to be noted. First, the study has proposed an influence on the daily returns, implying that the impact of such an influence during extreme events, such as a bank run, could be materially larger. We justify such a possibility on the grounds of specific behaviour in terms of possibly higher correlations during extreme events. The second point to note is that the tests have been performed with one type of web dataset. Given that there exists a vast number of such datasets, outcomes could be different in another test design. The tests presented have provided one approach capable of quantifiably justifying the propositions and therefore, have led us to understand the potential characteristics of OI and their integration in pricing models.

Our discovery has implications for an array of stakeholders. The conceptual basis defined for integrating web datasets, as alternative datasets in pricing models, provides a structural reasoning for the numerous studies in literature that have demonstrated their impact. For an investment decision-maker, OI is an avenue to mitigate their risks arising from incomplete information, if obtained from public and private sources only. Moreover, the demonstrated material significance implies a situation of *asymmetric information*, between those who integrate OI versus those who do not. This is therefore important for market quality, and presents a difficult regulatory situation. Given the web's open architecture and worldwide connectivity, existing forms of regulation for PI are not immediately applicable. Two important points to note in this regard are the identification of the source of information, and their spread across borders. An example is those social media platforms that operate worldwide, encouraging communication among network peers. Potentially, information about a domestically listed company might arise from one of the international peers, and spread across the web. When such information is material for the company, this poses a difficult regulatory situation. Regulatory bodies such as the SEC have been known to counter such occurrences on a case-to-case basis. Our discovery enables a stronger footing to analyse such cases uniformly.

[18] Results from the Diebold-Mariano test were statistically insignificant in most samples at 10%. However, when the p-values were compared for H0: Two methods have the same forecast accuracy, between the following alternative hypotheses, H1: AR(1)-X-GARCH is more accurate than AR(1)-GARCH, and H1: AR(1)-GARCH is more accurate than AR(1)-X-GARCH, the p-values obtained favoured the former.

[19] Results are expressed for one sample. Varying the sample changes the numeric values, whereas the order of percentage improvement remains similar.

To elaborate on the material significance from a management perspective, we explain that OI is integrable with investment decisions. From the DIKW[20] Pyramid [30], a dataset can be considered to be carrying information, and we refer to them as *informative datasets*. Such datasets are then provided as inputs to investment decisions, such as those modelled with TM.

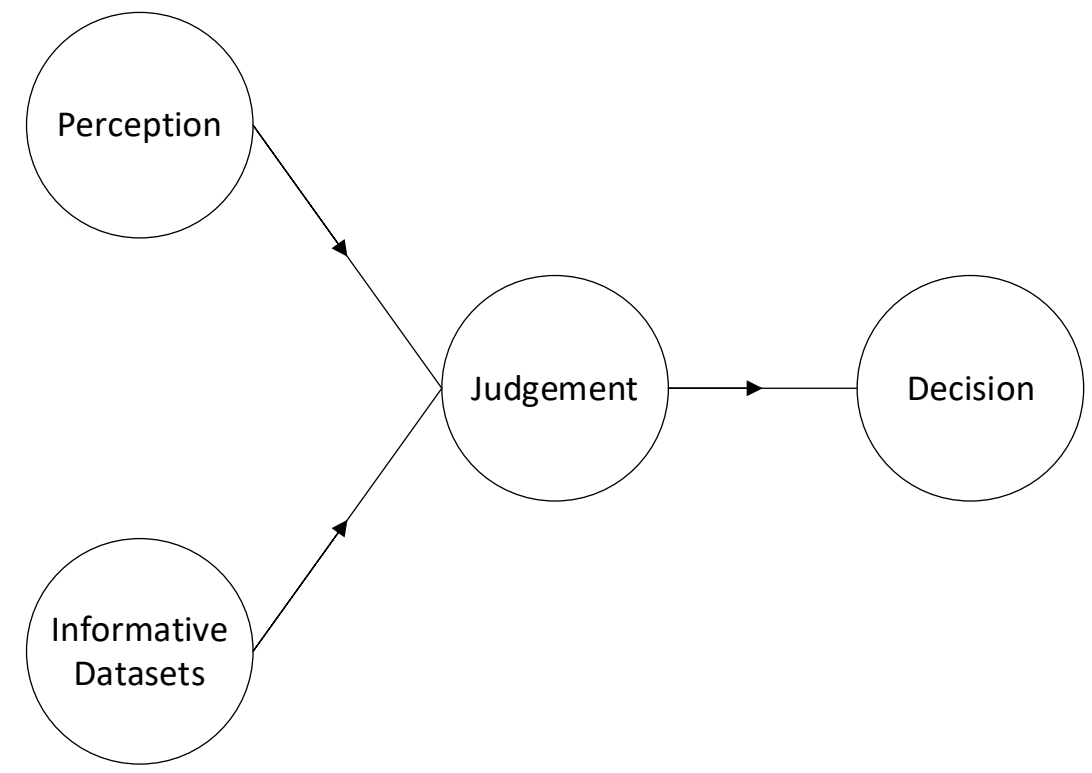

*Fig. 5. Informative datasets as input to investment decisions*

Fig. 5 extends the definition of TM to add informative datasets. Given that the literature on TM has modelled management decisions, such a definition allows addressing various aspects, such as cost estimates, analysing performance, estimating returns, and making data life-cycle choices. To stress on the costs involved, the characterisation of web datasets as OI implies that their integration is considered in terms of a *cost of information*, instead of a *cost of dataset*. This qualitative aspect influences accountability. The methodology we propose, therefore, allows the management to address a broad set of challenges while considering web datasets.

## VI. CONCLUSION

To summarise our research, we have proposed the discovery of OI arising from web datasets, identified their characteristics and provided a quantitative justification for propositions. Based on the test outcomes, we cannot rule out the material existence of OI. We have therefore provided a significant alternative to the notion that the web is just a faster way to disseminate public or private information. The discovery is significant in multiple ways. First, we provide a conceptual basis for the many studies that have demonstrated the impact of web datasets on pricing. Second, we have highlighted the differences between OI and PI notably through TABLE 1, therefore contributing to the understanding of the demarcation in their characteristics. Third, the testing framework is extensible to a web dataset and a price series of choice and therefore has the potential to integrate OI arising from a variety of sources. The impact of OI is important for an array of stakeholders, having the potential to explain significant market events such as a bank run or a flash crash. Furthermore, given the quantitative improvement in formulations, such an addition has the potential to improve performance of pricing models and consequently, investment decisions. Finally, the characteristic difference from PI is of importance from a regulatory standpoint, which might otherwise treat them through existing frameworks that are designed for public and private information sources only. Our research has therefore provided a new and defining perspective for information dissemination over the web.

[20] Data->Information->Knowledge->Wisdom

## ACKNOWLEDGMENTS

The research was performed as part of a Doctorate funded by the Centre of Excellence for Data Science, Artificial Intelligence and Modelling (DAIM) at the University of Hull. The size of the datasets necessitated use of high-performance computing resources. We acknowledge the Viper High Performance Computing facility of the University of Hull and its support team. We further acknowledge the support we have received from Dr Julius Mboli (University of Hull) and Professor Robert Hudson (University of Hull), whose observations during the progress have helped strengthen our research. A special note of thanks is attributed to Professor Kevin Pimbblet (University of Hull) who has pioneered new initiative with Data Science at University of Hull, thereby initiating this research.